\documentclass[twocolumn,showpacs,amsmath,amssymb,prd]{revtex4}
\usepackage{amsmath}
\usepackage{bm}
\usepackage{epsfig}
\usepackage[colorlinks=true,linkcolor=blue]{hyperref}

\newcommand{\be}{\begin{equation}}
\newcommand{\ee}{\end{equation}}

\newcommand{\bea}{\begin{eqnarray}}
\newcommand{\eea}{\end{eqnarray}}

\newcommand{\al}{\alpha}
\newcommand{\bt}{\beta}
\newcommand{\gm}{\gamma}

\newcommand{\dl}{\delta}
\newcommand{\Dl}{\Delta}
\newcommand{\eps}{\epsilon}

\newcommand{\lm}{\lambda}

\newcommand{\rh}{\rho}

\newcommand{\ta}{\tau}

\newcommand{\ph}{\phi}

\newcommand{\ch}{\chi}

\newcommand{\om}{\omega}
\newcommand{\Om}{\Omega}

\newcommand{\rarrow}{\rightarrow}
\newcommand{\Rarrow}{\Rightarrow}

\newcommand{\nn}{\nonumber}

\newcommand{\varep}{\varepsilon}

\begin{document}

\title {Graviton production in the scaling of a long-cosmic-string network}

\author{Kostas Kleidis$^1$, Apostolos Kuiroukidis$^2$, Demetrios B. Papadopoulos$^2$ and Enric Verdaguer$^3$}

\affiliation {$^1$Department of Mechanical Engineering, Technological Education Institute of Serres, GR-62124 Serres, Greece}

\affiliation {$^2$Department of Physics, Aristotle University of Thessaloniki, GR-54124 Thessaloniki, Greece}

\affiliation {$^3$Departament de Fisica Fonamental and Institut de Ciences del Cosmos, Universitat de Barcelona, Avinguda Diagonal 647, E-08028 Barcelona, Spain}

\date{\today}

\begin{abstract}
In a previous paper~\cite{1} we considered the possibility that (within the {\em early-radiation} epoch) there has been (also) a short period of a significant presence of cosmic strings. During this {\em radiation-plus-strings} stage the Universe matter-energy content can be modelled by a two-component fluid, consisting of radiation (dominant) and a cosmic-string fluid (subdominant). It was found that, during this stage, the cosmological gravitational waves (CGWs) - that had been produced in an earlier (inflationary) epoch - with comoving wave-numbers below a {\em critical} value (which depends on the physics of the cosmic-string network) were {\em filtered}, leading to a distorsion in the expected (scale-invariant) CGW {\em power spectrum}. In any case, the cosmological evolution gradually results in the {\em scaling} of any long-cosmic-string network and, hence, after a short time-interval, the Universe enters into the {\em late-radiation} era. However, along the transition from an early-radiation epoch to the late-radiation era through the radiation-plus-strings stage, the time-dependence of the cosmological scale factor is modified, something that leads to a discontinuous change of the corresponding scalar curvature, which, in turn, triggers the quantum-mechanical creation of {\em gravitons}. In this paper we discuss several aspects of such a process, and, in particular, the observational consequences on the expected gravitational-wave (GW) power spectrum.
\end{abstract}

\pacs{04.30.-w 11.25.-w 98.80.Cq}

\maketitle

\section{Introduction}

CGWs represent small-scale perturbations to the Universe metric tensor~\cite{2}. Since gravity is the weakest of the four known forces, these metric corrections decouple from the rest of the Universe at very early times, presumably at the Planck epoch~\cite{3}. Their subsequent propagation is governed by the space-time curvature~\cite{4}, encapsulating in the field equations the inherent coupling between relic GWs and the Universe matter content; the latter being responsible for the background gravitational field~\cite{5}.

In this context, we explore the creation of primordial GWs during the {\em scaling} of a long-cosmic-string network (see, e.g.,~\cite{6}), i.e., in the transition from a cosmological model in which the matter content can be modelled by a two-component fluid - consisting of radiation (dominant) and cosmic strings (subdominant) - to a {\em pure} radiation-dominated Universe.

Cosmic strings are one-dimensional objects that can be formed as {\em linear defects} at a symmetry-breaking phase transition~\cite{7},~\cite{8}. If they exist, they may help us to explain some of the large-scale structures seen in the Universe today, such as the gravitational lenses~\cite{9}. They may also serve as {\em seeds} for density perturbations~\cite{10},~\cite{11}, as well as potential sources of relic gravitational radiation~\cite{12},
\cite{13}.

The presence of cosmic strings in a radiation model, is responsible for the constancy of the {\em effective potential} which drives the temporal evolution of a CGW, leading to  a {\em critical value} $(k_c)$ of the comoving wave-number, which discriminates the metric perturbations into {\em oscillating} $(k > k_c)$ and {\em non-oscillating} $(k < k_c)$ modes~\cite{1}. As a consequence, the propagation of CGWs through a {\em radiation-plus-strings} stage would leave imprints on their {\em power spectrum}. However, such a stage does not last very long, since, gradually, the production of loops smaller than the horizon results in the {\em scaling} of the long-cosmic-string network. According to this process, the linear defects form a self-similar configuration, the density of which, eventually, behaves as $R^{-4}$ and the Universe (re)enters in the {\em pure-radiation} era (see, e.g.,~\cite{7}).

In the present article, we study the quantum-mechanical production of {\em gravitons} in the transition of the Universe from an {\em early-radiation} epoch to the {\em late-radiation} era through the radiation-plus-strings stage. Several theoretical (and observational) consequences are discussed.

The paper is organized as follows: In Sec. II, we summarize the theory of CGWs propagating in curved space-time, including the implications arising from the constancy of the effective potential. In Sec. III, we take advantage of the scaling of a long-cosmic-string network to arrive at the late-radiation era. In Sec. IV, we demonstrate that, due to the modification in the time-dependence of the cosmological scale factor along this transition, the scalar curvature changes discontinuously; hence, gravitons are created, and, in Sec. V, we explore the corresponding {\em power spectrum}. The high-frequency part $(k > k_c)$ of this spectrum acquires a {\em characteristic profile}, resulting in a {\em periodic function of the frequency}. Finally, in Sec. VI, we evaluate the (integrated) energy density of the CGWs created in the scaling of the long-cosmic-string network and we compare it with the corresponding quantity predicted by inflation (see, e.g.,~\cite{12}). It is worth noting that, the energy density of the CGWs created in the transition of the Universe from an early-radiation epoch to the late-radiation era through a radiation-plus-strings stage, not only is added to the corresponding inflationary quantity. In fact, at times $10^{-19} \; sec \leq t \leq 6540 \; sec$, i.e., within the first (and, certainly, most interesting) $109 \; min$ of the Universal evolution, it dominates over every other form of GW-energy density.

As regards the space-time geometry itself, we confine ourselves to the spatially flat Friedmann - Robertson - Walker (FRW) model. This model appears to interpret adequately both the observational data related to the known thermal history of the Universe and the theoretical approach to cosmic-string configurations~\cite{7}. Therefore, it seems to be the most appropriate candidate for the curved background needed for this study.

\section{CGWs in a Friedmann model}

The quantum-gravitational creation of gravitons in an expanding FRW Universe was first demonstrated by Grishchuk~\cite{14}. He showed that, in the linear approximation, the behavior of a CGW propagating in curved space-time is identical to that of a massless, minimally-coupled scalar field. In particular, each of the two polarization states of the metric perturbation satisfy the Klein-Gordon equation \be h_{ij \: ; \al}^{\; \; \; \: ; \al} = 0 , \ee where Greek indices refer to the four-dimensional space-time, Latin indices refer to the corresponding three-dimensional spatial slices and the {\em semicolon} denotes covariant derivative. The quantization of primordial GWs and that of the minimally-coupled, massless scalar fields also proceeds along identical lines, as it was demonstrated by Ford and Parker~\cite{15},~\cite{16}.

A weak CGW $(\vert h_{ij} \vert \ll 1)$ propagating in a spatially-flat FRW cosmological model is defined as~\cite{12} \be ds^2 = c^2 dt^2 - R^2(t) \left ( \dl_{ij} + h_{ij} \right ) dx^i dx^j , \ee where $c$ is the velocity of light, $\dl_{ij}$ is the Kronecker symbol and the dimensionless scale factor $R(t)$ is a solution to the Friedmann equations with matter content in the form of a {\em perfect fluid}. In terms of the {\em conformal time}, \be \ta = \int \frac{dt}{R(t)}, \ee the solution to the Klein-Gordon equation (1) is a {\em linear superposition of plane-wave modes} \be h_{ij} (\ta , x^r) = \phi_k (\ta) \: \varep_{ij} \: e^{\imath k_r x^r} , \ee where $\phi_k (\ta)$ is a (complex) function of time and $\varep_{ij}$ is the polarization tensor, depending only on the direction of the {\em comoving} wave-vector $k_r$. For a fixed wave-number $k^2 = \sum k_r^2$, the time-dependent part of the corresponding GW mode satisfies the second-order differential equation (see, e.g.,~\cite{17}) \be \phi_k^{\prime \prime} + 2 \frac{R^{\prime}}{R} \: \phi_k^{\prime} + k^2 c^2 \: \phi_k = 0 , \ee where a {\em prime} denotes differentiation with respect to $\ta$. Under the further decomposition \be h_k (\ta) = \phi_k (\ta) \: R (\ta), \ee Eq. (5), which governs the temporal evolution of a CGW in a FRW Universe, is written in the form \be h_k^{\prime \prime} + \left ( k^2 c^2 - \frac{R^{\prime \prime}}{R} \right ) \: h_k = 0 . \ee Eq. (7) can be treated as the Schr\"{o}dinger equation, for a particle moving in the presence of the {\em effective potential} \be V_{eff} = \frac{R^{\prime \prime}}{R} . \ee To solve this equation, one needs an evolution formula for the cosmological model under consideration.

In terms of the conformal time, the spatially-flat FRW model is a solution to the Friedmann equation \be \left ( \frac{R^{\prime}}{R^2} \right )^2 = \frac{8 \pi G}{3} \; \rh (\ta) \ee (where $G$ is Newton's universal constant of gravitation), with matter content in the form of a perfect fluid source, $T_{\mu \nu} = diag (\rho c^2, - p, -p, -p)$, which obeys the conservation law \be \rh^{\prime} + 3 \frac{R^{\prime}}{R} \; \left ( \rh + \frac{1}{c^2} \; p \right ) = 0 \ee and the equation of state \be p = \left ( \frac{m}{3} - 1 \right ) \; \rh c^2 , \ee where $\rh(\ta)$ and $p(\ta)$ represent the mass density and the pressure, respectively.

The linear equation of state (11) covers most of the matter components considered to drive the evolution of the Universe, such as {\em quantum vacuum} $(m = 0)$, a network of {\em domain walls} $(m = 1)$, a gas of {\em cosmic strings} $(m = 2)$, {\em dust} $(m = 3)$, {\em radiation} $(m = 4)$ and Zel'dovich {\em ultra-stiff matter} $(m = 6)$. For each component, the continuity equation (10) yields \be \rh = \frac{M_m}{R^m} , \ee where $M_m$ is an integration constant. Provided that the various components do not interact with each other, a mixture of them obeys~\cite{18} \be \rh = \sum_m \frac{M_m}{R^m} , \ee where, now, Eq. (10) holds for every matter constituent separately.

\subsection{Constancy of the effective potential}

There is a case of particular interest, involved in the temporal evolution of a primordial CGW, in which the effective potential is {\em constant} for every $\ta$, namely \be \frac{R^{\prime \prime}}{R} = const = \frac{8 \pi G}{3} M_2 , \ee where $M_2$ is a non-negative constant of units $gr/cm^3$ \cite{1}. In this case, Eq. (7) is written in the form \be h_k^{\prime \prime} + \left ( k^2 c^2 - \frac{8 \pi G}{3} M_2 \right ) h_k = 0 . \ee From Eq. (15), it becomes evident that, as long as $M_2 \neq 0$, there exists a {\em critical value} of the comoving wave-number, \be k_c = \sqrt{\frac{8 \pi G}{3 c^2} M_2 } \: , \ee which discriminates the primordial GWs into {\em oscillating} ($k > k_c$) and {\em non-oscillating} ($k < k_c$) modes. According to~\cite{1}, such a property is met in a cosmological model with matter content in the form of a two-component fluid with \be \rh (\ta) = \frac{M_2}{R^2} + \frac{M_4}{R^4} \: , \ee where $M_4$ is an integration constant, also of units $gr/cm^3$. 

For $M_4 = 0$, Eq. (17) results in $\rh \sim R^{-2}$, which, by virtue of Eqs. (11) and (12), leads to $p = - \frac{1}{3} \rh c^2$. This case, now known as the {\em string-dominated Universe}~\cite{19}, was considered by Ford and Parker~\cite{15}, in their effort for a natural {\em renormalization} of the energy-momentum tensor of the gravitational-perturbation field. 

A string-dominated Universe does not seem likely and, therefore, in this article, we consider $M_4 \neq 0$. Accordingly, upon consideration of Eqs. (12) and (13), we distinguish two cases, in Eq. (17), with respect to $M_2$: (i) $M_2 = 0$ and (ii) $M_2 \neq 0$. We consider each one of them separately.

\subsection{CGWs in a radiation-dominated Universe}

For $M_2 = 0$, Eq. (17) reads \be \rh (\ta) = \frac{M_4}{R^4}. \ee In view of Eq. (12), this is the matter-energy content of a Universe filled with relativistic particles - radiation. Now, the effective potential vanishes and Eq. (15) is written in the form \be h_k^{\prime \prime} + k^2 c^2 h_k = 0, \ee i.e., it is a simple harmonic oscillator equation; like a wave propagating in flat space-time, where, there is no particle production~\cite{20}. Therefore, the choice of a solution to Eq. (19) corresponds to the choice of an {\em initial (or final) quantum state} for the gravitational-perturbation field. Eq. (19) admits the solution \be h_k^{in} (\ta) = \sqrt {\frac{2}{\pi k c}} \: e^{-\imath k c \ta} \ee and the gravitational-perturbation field is in the well-defined {\em adiabatic-vacuum} state~\cite{21}.

\subsection{CGWs in the presence of cosmic strings}

Setting, in Eq. (17), $M_2 \neq 0$, corresponds to a radiation model {\em contaminated} by cosmic strings, in which, the two constituents (relativistic particles and linear defects) do not interact with each other (cf. Eq. (13)). In this case, the constant effective potential (14), is related to the initial amount of the linear defects in the {\em "mixture"}, $M_2$. Now, Eq. (15) results in \be h_k^{\prime \prime} + \left ( k^2 - k_c^2 \right ) c^2 \: h_k = 0 \: . \ee For every $0 < k < \infty$, the general solution to Eq. (21) can be expressed in the form \be h_k (\ta) = \sqrt {\frac{2}{\pi \gm k c}} \: \left [ c_1 \: e^{- \imath \gm k c \ta} + c_2 e^{+ \imath \gm k c \ta} \right ] , \ee where $c_1$ and $c_2$ are complex numbers, satisfying the {\em Wronskian} condition \be \vert c_1 \vert^2 - \vert c_2 \vert^2 = 1 \ee (as a consequence of the mode normalization when the gravitational perturbations are quantized), and we have introduced the parameter \be \gm = \left [ 1 - \left ( \frac{k_c}{k} \right )^2 \right ]^{1/2} , \ee which measures the departure from the {\em pure-radiation} case, given by Eq. (20). 

\section{A tale of cosmic strings}

The presence of cosmic strings in the Universe is purely a question of topology~\cite{22}. In particular, after inflation (and reheating) the Universe enters in an {\em early-radiation} (ER) epoch~\cite{23}, during which, the temperature drops monotonically $(T \sim R^{-1})$. This cooling process may
have resulted in the breaking of a fundamental U(1) {\em local gauge symmetry}, which, in turn, led to the formation of {\em linear topological defects} (for a detailed analysis see \cite{7} and/or \cite{8}). Within this era, the cosmological scale factor behaves as \be R_{ER} (t) = \left ( \frac{t}{t_{cr}} \right )^{1/2} \; \; \Rarrow \; R_{ER} (\ta) = \frac{\ta}{\ta_{cr}} , \ee where $\ta_{cr} = 2 \: t_{cr}$ marks the (conformal) time at which the cosmic strings are formed, and we have normalized $R (t_{cr})$ to unity. 

By the time these linear defects are formed, they are moving in a very dense environment and, hence, their motion is heavily damped, due to {\em string-particle scattering}~\cite{24},~\cite{25},~\cite{26}. This {\em friction} becomes subdominant to expansion damping~\cite{24} at \be t_* = \left ( \frac{G \mu}{c^2} \right )^{-1} \; t_{cr} \; \; \Rarrow \; \ta_* = \left ( \frac{G \mu}{c^2} \right )^{-1/2} \; \ta_{cr} , \ee where $\mu$ is the {\em mass per unit-length} of the linear defect. For $\ta \geq \ta_*$, the motion of the long cosmic strings is essentially independent of anything else in the Universe and soon they acquire relativistic velocities~\cite{7}. Now, the {\em gas} of cosmic strings does not interact with the fluid of relativistic particles and, hence, Eq. (17) holds. Therefore, we may consider that, for $\ta \geq \ta_*$, the evolution of the curved space-time is driven by a {\em two-component fluid}, consisting of radiation (dominant) and cosmic strings (subdominant). Accordingly, $\ta_*$ marks the beginning of a {\em radiation-plus-strings} stage, during which, the Friedmann equation (9) admits the solution~\cite{1} \be R (\ta) = \sqrt{\frac{M_4}{M_2}} \; \sinh \sqrt{\frac{8 \pi G}{3} M_2} \; \ta . \ee Nevertheless, the scale factor (27) can drive the Universe expansion only for a short period of time after $\ta_*$, since, cosmic strings should (at any time) be a small proportion of the Universe matter-energy content. This means that the equation of state (17) has validity only for a limited time-period, otherwise, cosmic strings would, eventually, dominate on the Universe energy density (see, e.g.,~\cite{19}).

In fact, the radiation-plus-strings stage (if ever existed) does not last very long. Numerical simulations at the early 90's~\cite{27} -~\cite{30}, as well as their present-time counterparts~\cite{31} -~\cite{33}, suggest that, after the friction becomes unimportant, the production of loops smaller than the {\em horizon}, gradually, results in the {\em scaling} of the long-cosmic-string network. According to this process, the linear defects form a {\em self-similar configuration}, the density of which, eventually, behaves as $R^{-4}$. In this way, at some (physical) time $t_{sc} > t_*$ the Universe re-enters in the ({\em late}) radiation (LR) era, \be R_{LR} (t) = R_{sc} \: \left ( \frac{t}{t_{sc}} \right )^{1/2} \; \; \Rarrow \; R_{LR} (\ta) = R_{sc} \: \frac{\ta}{\ta_{sc}} , \ee before it can become {\em string-dominated}. The duration of the radiation-plus-strings stage is quite uncertain, mostly due to the fact that numerical simulations (which revealed the scaling of the long-strings network) can be run for relatively limited times. For example, the earliest treatments~\cite{29},~\cite{30} suggested that $t_{sc} \simeq 30 \: t_*$ $(\ta_{sc} \simeq 5.5 \: \ta_*)$, while, the most recent ones~\cite{32} raise this value to $t_{sc} \simeq 300 \: t_*$ $(\ta_{sc} \simeq 17 \: \ta_*)$. According to Eq. (40) of ~\cite{1}, the former result leads to $k_c c \Dl \ta = k_c c (\ta_{sc} - \ta_*) = 0.09$, while, adoption of the latter result, would lead to $k_c c \Dl \ta = 0.32$.

\section{Graviton creation}

Around $\ta_*$ and $\ta_{sc}$, both $R (\ta)$ and $R^{\prime} (\ta)$ can be matched to be continuous (cf.~\cite{1}), but $R^{\prime \prime} (\ta)$ is {\em essentially discontinuous}, acquiring a {\em non-zero value} through Eq. (14). Consequently, the scalar curvature of the space-time changes discontinuously both at $\ta_*$ and at $\ta_{sc}$. A discontinuous change in the scalar curvature {\em produces gravitons}~\cite{20},~\cite{34}.

As a consequence, the evolution of the CGW modes in the transition of the Universe from an early-radiation epoch to the late-radiation era through a radiation-plus-strings stage, is modified as follows:

\begin{itemize}

\item For $\ta \leq \ta_*$ the effective potential (14) vanishes. The temporal evolution of the metric perturbations is governed by Eq. (19), admitting the solution (20) and the gravitational-perturbation field is in an adiabatic-vacuum state.

\item For $\ta_* < \ta < \ta_{sc}$ the effective potential reduces to a non-vanishing constant $(V_{eff} = k_c^2 c^2)$ and the evolution of the metric perturbations is driven by Eq. (21). Due to the discontinuous change of the scalar curvature at $\ta_*$, during the radiation-plus-strings stage the gravitational-perturbation field is no longer in its vacuum state and the general solution to Eq. (21) is the linear superposition of positive- and negative-frequency solutions (22). Moreover, for every $\ta_* < \ta < \ta_{sc}$, the primordial CGWs, being discriminated by the value of their comoving wave-number (in comparison to $k_c$), do not evolve in a unique way: The GW modes with $k > k_c$ {\em perform oscillations} within the {\em horizon}, while, those of $k < k_c$ remain outside of the {\em Hubble sphere} and {\em grow exponentially}~\cite{1}.

\item Finally, for $\ta \geq \ta_{sc}$, $V_{eff} = 0$ and the temporal evolution of a metric perturbation in curved space-time is (once again) governed by Eq. (19). However, due to the discontinuity of the scalar curvature at $\ta_{sc}$, this time, the general solution to Eq. (19) is written in the form \be h_k^{out} (\ta) = \sqrt{\frac{2}{\pi k c}} \; \left ( \al_k \: e^{- \imath k c \ta} + \bt_k \: e^{+ \imath k c \ta} \right ) , \ee where the ({\em Bogolubov}) coefficients $\al_k$ and $\bt_k$ should satisfy the Wronskian condition \be \vert \al_k \vert^2 - \vert \bt_k \vert^2 = 1 . \ee 

\end{itemize}

Eq. (29) determines the {\em final quantum state} of the gravitational-perturbation field. As a result of {\em conformal invariance}, this state is also an adiabatic vacuum, but, as long as $\bt_k \neq 0$, it differs from the corresponding state of the early-radiation epoch. Clearly, the occurrence of a non-zero $\bt_k$ in the late-radiation era, would signal that $h_k^{out} (\ta)$ is not a {\em pure} positive-frequency solution, but contains also a negative-frequency component. This means that, if the {\em in-state} of the gravitational-perturbation field is taken to be vacuum, then, {\em particles} are found in the {\em out-state}~\cite{21}.

In what follows, we discuss the evolution of the modes (20), both with $k > k_c$ and with $k < k_c$, in the transition of the Universe from the early-radiation epoch to the late-radiation era, through a radiation-plus-strings stage. The coefficients $c_1$, $c_2$, $\al_k$ and $\bt_k$ can be determined by the requirement that $h_k (\ta)$ and its first derivative $h_k^{\prime} (\ta)$ are continuous across the boundaries $\ta_*$ and $\ta_{sc}$.

Matching the modes (20) and (22), as well as their first derivatives, at $\ta = \ta_*$, we obtain \be c_1 = \frac{\gm + 1}{2 \sqrt {\gm}} \; e^{\imath k c (\gm - 1) \ta_*} \ee and \be c_2 = \frac{\gm - 1}{2 \sqrt{\gm}} \; e^{- \imath k c (\gm + 1) \ta_*} , \ee for which, the Wronskian condition (23) holds. Similarly, matching the modes (22) and (29), as well as their first derivatives, at $\ta = \ta_{sc}$, we obtain \be \al_k = \frac{1}{4 \gm} \: \left [ (\gm + 1)^2 \: e^{- \imath (\gm - 1) k c \Dl \ta} - (\gm - 1)^2 \: e^{+ \imath (\gm + 1) k c \Dl \ta} \right ] \ee and \be \bt_k = \imath \: \frac{\gm^2 - 1}{2 \gm} \: \sin (\gm k c \Dl \ta) \: e^{- \imath k c (\ta_* + \ta_{sc})} , \ee where $\Dl \ta = \ta_{sc} - \ta_*$ and we have taken into account Eqs. (31) and (32). By virtue of Eq. (34), {\em the number of gravitons}~\cite{21} created in the mode denoted by $0 < k < \infty$, is written in the form \be N_k = \vert \bt_k \vert^2 = \frac{1}{4 \gm^2} \left ( \gm^2 - 1 \right )^2 \: \sin^2 ( \gm k c \Dl \ta ) \ee and the normalization condition $\vert \al_k \vert^2 - \vert \bt_k \vert^2 = 1$, holds. The {\em observable quantity} $N_k$ possesses a series of very interesting properties, depending on the value of $k$ in comparison to $k_c$. We distinguish the following cases:

\subsection{Oscillating modes}

In this case, $k > k_c$ and Eq. (35) yields \be N_{k > k_c} = \frac{k_c^4}{4 \: k^2} \: \frac{\sin^2 \left ( c \Dl \ta \sqrt{k^2 - k_c^2} \right )}{k^2 - k_c^2} \: . \ee Obviously, for $k_c = 0$, $N_{k > k_c}$ vanishes, i.e., no gravitons are created in the absence of cosmic strings.

On the other hand, for $k \rarrow k_c$ and/or $\Dl \ta \rarrow 0$, Eq. (36) results in \be N_{k > k_c} \rarrow \frac{k_c^4}{4 \: k^2} \; c^2\: \Dl \ta^2 . \ee Now, we distinguish the following cases:

\begin{enumerate}

\item[(i)] $\Dl \ta = 0$: In this case, $N_{k > k_c} = 0$. Although this may look like an unexpected result, it is not. Usually, the instantaneous transition from one epoch to another may lead to pathological results, especially when high-frequency modes are concerned (see, e.g.,~\cite{21}). However, in our case, as long as $\Dl \ta = 0$ no transition ever takes place. The Universe remains in the (early) radiation era, where {\em no gravitons are produced}.

\item[(ii)] $k = k_c$: In this case, the number of gravitons created in the scaling of a long-cosmic-string network acquires the constant value \be N_{k_c} = \frac{k_c^2}{4} \: c^2 \: \Dl \ta^2 . \ee Eq. (38) suggests that, the number of gravitons in the {\em ground state} $(k_c)$ of the oscillating modes is well-defined and finite. 

\item[(iii)] Finally, for $k \gg k_c$ and $k \rarrow \infty$, $N_{k > k_c} \rarrow 0$ and particle production ceases. Consequently, there should be no {\em ultraviolet divergences}.

\end{enumerate}

On the other hand, according to Eq. (36), the {\em expectation value} of the {\em number operator}, $N_{k > k_c}$, varies periodically with $k$. Hence:

\begin{itemize}

\item For comoving wave-numbers satisfying the condition \be c \: \Dl \ta \sqrt{k^2 - k_c^2} = m \pi \; \; (m = 1, 2, ...) , \ee $N_{k > k_c} = 0$, admitting its absolute minimum value. The case $m = 0$ is excluded due to the fact that $N_{k_c} \neq 0$. As a consequence, creation of gravitons with comoving wave-numbers \be k_m^2 = k_c^2 + m^2 \: \frac{\pi^2}{(c \Dl \ta)^2} \; \; (m = 1, 2, ...) , \ee never occurs by this mechanism, a case which (in quantum mechanics) is known as {\em perfect transmition} (see, e.g.,~\cite{35}, pp. 101 - 104). Clearly, the spectrum of these states is {\em discrete} and their position on the $k$-axis is determined (solely) by the choice of $\Dl \ta \neq 0$.

\item On the other hand, the values of $k > k_c$ that label the maxima of $N_{k > k_c}$ obey the equation \be \left (c \Dl \ta \sqrt {k^2 - k_c^2} \right ) \cot \left ( c \Dl \ta \sqrt{k^2 - k_c^2} \right ) = 2 - \left ( \frac{k_c}{k} \right )^2, \ee which can be  solved {\em numerically}, to give \be k_n^2 \simeq k_c^2 + \left (n + \frac{49}{36} \right )^2 \: \frac{\pi^2}{(c \Dl \ta)^2} \; \; (n = 0, 1, 2, ...) . \ee Once again, the position of these (discrete) states on the $k$-axis is determined (solely) by the choice of $\Dl \ta \neq 0$. Therefore, a possible detection of these waves, among other things, could reveal valuable information on the duration of the radiation-plus-strings stage, as well. 

\end{itemize}

\subsection{Non-oscillating modes}

According to~\cite{1}, for $\ta_* < \ta < \ta_{sc}$, the CGW modes with $k < k_c$ remain outside the horizon, where they grow exponentially. Nevertheless, from the quantum-mechanical point of view, low-frequency (non-oscillatory) modes are also created along the transition of the Universe through the radiation-plus-strings stage, in the form of waves that {\em tunnel} through the {\em barrier} $V_{eff}$ of {\em width} $\Dl \ta$. Of course, doing so, they are subject to an additional exponential decrease (see, e.g.,~\cite{21}), which, eventually, may counter-balance their original growth. Nevertheless, due to the {\em tunneling effect}, they can also contribute to the CGWs with $k > k_c$ created during the radiation-plus-strings stage, and, therefore, they should be taken into account in the late-radiation era. 

For $k < k_c$, the expectation value (35) of the number-operator, $N_{k < k_c} = \vert \bt_{k < k_c} \vert^2$, is given by \be N_{k < k_c} = \frac{k_c^4}{4 \: k^2} \: \frac{\sinh^2 \left ( c \Dl \ta \sqrt{k_c^2 - k^2} \right )}{k_c^2 - k^2} \: . \ee Once again, either for $k_c = 0$ or for $\Dl \ta = 0$, $N_{k< k_c}$ vanishes, i.e., no tunneling ever takes place. On the other hand, for $k \rarrow k_c$, Eq. (43) is reduced to the constant value (38), verifying that, the number of gravitons in the state denoted by $k_c$ is well-defined and finite.

In view of Eqs. (36), (38) and (43), it is now evident that, the number operator involved in the creation of gravitons along the transition of the Universe through a radiation-plus-strings stage is a {\em continuous function} of $k$, for every $0 < k < \infty$.

Finally, for $k \ll k_c$, the expectation value of $N_{k < k_c}$ reads \be N_{k < k_c} \simeq \frac{k_c^2}{4 k^2} \: \sinh^2 (k_c c \Dl \ta) . \ee The apparent {\em infrared divergence} in Eq. (44) is {\em fictitious}. Indeed, these very low-frequency (very long-wavelength) waves are created way beyond the horizon~\cite{1}. As a consequence, they will enter inside the visible Universe at very late times, some of them (i.e., those with $k \approx 0$) not even today. 

\section{The power spectrum of the created gravitons}

During the (late) radiation stage of the cosmic expansion, the {\em physical} frequency of a GW is defined as $\om = c k / R(\ta)$ and the number of states with frequencies in the interval $\om$ and $\om + d \om$ is written in the form $d n_{\om} = \om^2 d \om / 2 \pi^2 c^3$. Accordingly, the distribution of the gravitational-energy density, summed over the two polarization states of the gravitons present in the late-radiation epoch, is given by \be d \eps_{gw} = {\cal P} (\om) d \om = 2 \hbar \om \: \frac{\om^2}{2 \pi^2 c^3} \: \vert \bt_k \vert^2 \: d \om \: . \ee Once again, we distinguish the following cases:

\subsection{Oscillating modes}

In terms of $k > k_c$, Eq. (45) is written in the form \bea &&d \eps_{k > k_c} = {\cal P} (k > k_c) d k = \\ && \frac{1}{R^4 (\ta)} \: \frac{\hbar c}{4 \pi^2} \: k_c^4 \: \frac{\sin^2 \left ( c \Dl \ta \sqrt{k^2 - k_c^2} \right )}{k^2 - k_c^2} \: k \: d k . \nn \eea Now, in view of Eq. (46), the corresponding {\em power spectrum}~\cite{17} reads \bea {\cal P} (k > k_c) & = & \frac{d \eps_{k > k_c}}{d k} = \\ & = & \frac{\hbar c}{4 \pi^2} \: k_{c_{ph}}^4 (\ta) \: \frac{\sin^2 \left ( c \Dl \ta \sqrt{k^2 - k_c^2} \right )}{k^2 - k_c^2} \: k  \: , \nn \eea where we have introduced the (physical) critical wave-number at every $\ta$, as \be k_{c_{ph}} (\ta) = k_c / R (\ta). \ee For $k$ close to $k_c$, ${\cal P} (k > k_c) \sim k$, while, in the large-$k$ limit, it results in a {\em damped oscillation} \be {\cal P} (k > k_c) \sim \frac{1}{k} \; \sin^2 (k c \Dl \ta) . \ee According to Eq. (49), {\em the production of gravitons with $k > k_c$ is suppressed at high frequencies}.

Similarly, the {\em logarithmic spectrum} is written in the form \bea \frac{d \eps_{k > k_c}}{d (\ln k)} & = & k {\cal P} (k > k_c) = \\ & = & \frac{\hbar c}{4 \pi^2} \: k_{c_{ph}}^4 (\ta) \: \frac{\sin^2 \left ( c \Dl \ta \sqrt{k^2 - k_c^2} \right )}{k^2 - k_c^2} \: k^2 \nn . \eea Eq. (50) becomes particularly {\em transparent} (and useful) in terms of the dimensionless parameter (see, e.g.,~\cite{1}) \be x = \frac{k}{k_c} \geq 1 , \ee which measures the comoving wave-number of the oscillating modes in units of their {\em ground-state} counterpart, $k_c$. In terms of $x$, Eq. (50) is written in the form \bea \frac{d \eps_{k > k_c}}{d (\ln k)} & = & k {\cal P} (k > k_c) = x {\cal P} (x > 1) \\ & = & \frac{\hbar c}{4 \pi^2} \: k_{c_{ph}}^4 (\ta) \: \frac{\sin^2 \left ( k_c c \Dl \ta \sqrt{x^2 - 1} \right )}{x^2 - 1} \: x^2 , \nn \eea or, else, \bea x {\cal P} (x > 1) & = & 4 \pi \; \frac{1}{4} \left ( k_c \: c \: \Dl \ta \right )^2 \; 2 \hbar \: c \: k_{c_{ph}} (\ta) \; \frac{1}{\lm_{c_{ph}}^3 (\ta)} \nn \\ & \times & \frac{\sin^2 \left ( k_c c \Dl \ta \sqrt{x^2 - 1} \right )}{\left ( k_c c \Dl \ta \sqrt{x^2 - 1} \right )^2} \; x^2 , \eea where $\lm_{c_{ph}} = 2 \pi / k_{c_{ph}}$ is the {\em correlation (wave)length}, associated to $k_{c_{ph}}$.

\begin{figure}[ht!]
\centerline{\mbox {\epsfxsize=9.cm \epsfysize=7.5cm \epsfbox{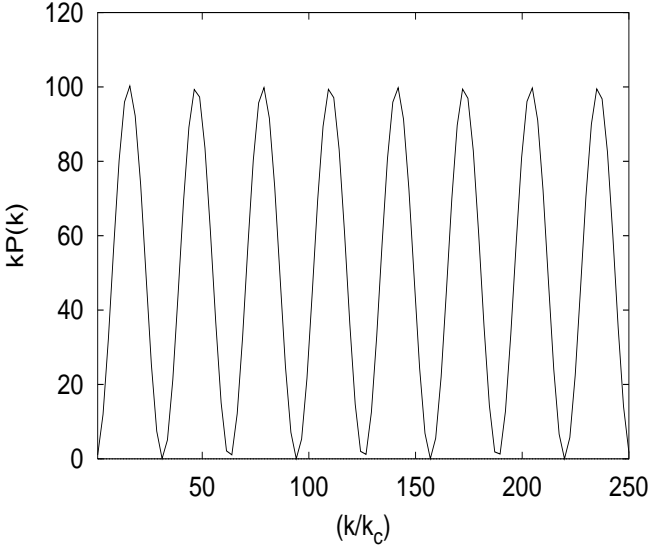}}} \caption{The logarithmic spectrum $k \: {\cal P} (k)$, of the high-frequency part $(k > k_c)$ of the metric perturbations created in the scaling of a long-cosmic-string network (normalized over $4 \pi \: \eps_c$), versus $k$ in units of $k_c$, for $\Dl \ta = 5.5 \: \ta_*$ (according to~\cite{1}).}
\end{figure}

At a fixed $\ta \geq \ta_{sc}$, the gravitational energy contained within the volume $\lm_{c_{ph}}^3$, is $2 \hbar c k_{c_{ph}}$, i.e., the energy of a graviton with physical wave-number $k_{c_{ph}}$ summed over the two polarization states. Therefore, the quantity \be \eps_c (\ta) = N_{k_c} \; 2 \hbar \: c \: k_{c_{ph}} (\ta) \; \frac{1}{\lm_{c_{ph}}^3 (\ta)} \ee (where $N_{k_c}$ is given by Eq. (38)) represents the energy density of the gravitons created in the (ground) state denoted by $k_c$, at every $\ta \geq \ta_{sc}$ and the logarithmic spectrum (53) reads \be x {\cal P} (x > 1) = 4 \pi \: \eps_c (\ta) \: \frac{\sin^2 \left ( k_c c \Dl \ta \sqrt{x^2 - 1} \right )}{\left ( k_c c \Dl \ta \sqrt{x^2 - 1} \right )^2} \: x^2 . \ee We can easily verify that, for $x = 1$, one is left with the energy density of the gravitons in the state $k = k_c$ (within the solid angle of a sphere - $4 \pi$), while, as we depart from the lowest-allowed comoving wave-number of the oscillating modes, the gravitational power is distributed among the various $k$-intervals according to Eq. (55). In particular, on the approach to $k_c$, the logarithmic spectrum is {\em quadratic in $k$}, while, for $k \gg k_c$ $(x \rarrow \infty)$, it reduces to a periodic function of the comoving wave-number with {\em constant amplitude} (Fig. 1) \be \frac{d \eps_{k > k_c}}{d (\ln k)} \sim \sin^2 (k c \Dl \ta) . \ee 

The importance of this result rests in the fact that the logarithmic spectrum is a re-scaling of the function \be \frac{1}{\varep_{cr}} \: \frac{d \eps_{gw}}{d (\ln k)} = \Om_{gw} = \frac{1} {\varep_{cr}} \: \frac{d \eps_{gw}}{d (\ln f)} \ee (where $\varep_{cr}$ is the {\em critical-energy density for closing the Universe}), which characterizes the {\em intensity} of a {\em stochastic background} of CGWs~\cite{36}. 

Therefore, if a relic CGW signal has been {\em filtered} by a long-cosmic-string network~\cite{1}, its {\em profile} would have been modified along the lines of Eq. (55) (or Eq. (56)), thus, resulting in a periodic function of the frequency, the period of which depends solely on the duration of the corresponding radiation-plus-strings stage (Fig. 1).

\subsection{Non-oscillating modes}

In this case, in terms of the dimensionless parameter $x = \frac{k}{k_c} \: (\leq 1)$, the distribution of the gravitational-energy density of the waves with $k < k_c$, which are present in the late-radiation epoch, is given by \be d \eps_{k < k_c} = 4 \pi \: \eps_c (\ta) \: \frac{\sinh^2 \left ( k_c c \Dl \ta \sqrt{1 - x^2} \right )}{\left ( k_c c \Dl \ta \sqrt{1 - x^2} \right )^2} \: x \: dx, \ee where $\eps_c (\ta)$ is given by Eq. (54). Accordingly, the corresponding logarithmic spectrum reads (notice that, here, ${\cal P}_k$ stands for $P_k^2$ of~\cite{1}) \bea \frac{d \eps_{k < k_c}}{d (\ln k)} & = & k {\cal P} (k < k_c) = \\ & = & 4 \pi \: \eps_c (\ta) \: \frac{\sinh^2 \left ( k_c c \Dl \ta \sqrt{1 - x^2} \right )}{\left ( k_c c \Dl \ta \sqrt{1 - x^2} \right )^2} \: x^2 \nn \eea and is presented in Fig. 2, which, in fact, corresponds to a (magnified) {\em "detail"} of the low-left corner of Fig. 1.

\begin{figure}[ht!]
\centerline{\mbox {\epsfxsize=9.cm \epsfysize=7.5cm \epsfbox{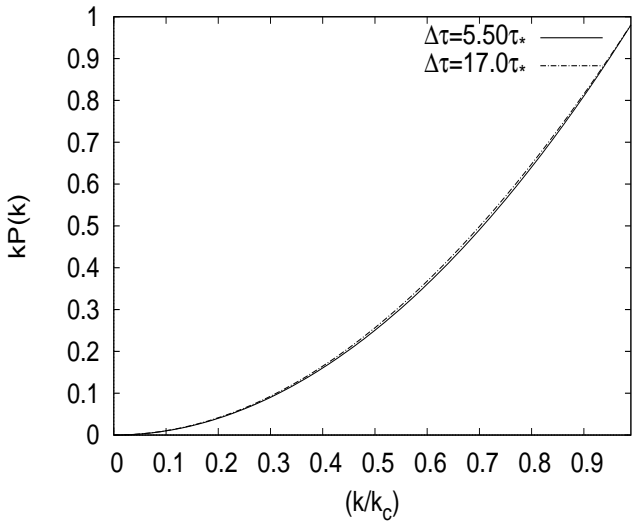}}} \caption{The logarithmic spectrum $k \: {\cal P} (k)$, of the CGW modes with $k < k_c$, i.e., those created outside the horizon (normalized over $4 \pi \: \eps_c$), versus $k$ in units of $k_c$, for $\Dl \ta = 5.5 \: \ta_*$ (solid line) and $\Dl \ta = 17 \: \ta_*$ (dashed line).}
\end{figure}

Accordingly, if the Universal evolution has undergone through a radiation-plus-strings stage, then, in connection to~\cite{1}, the {\em anticipated profile} of a stochastic GW signal should be modified as follows:

\begin{itemize}

\item Every GW background of cosmological origin that created earlier than $\ta_*$ (e.g., the metric perturbations predicted by inflation), is being {\em filtered} by the long-cosmic-string network, thus, being discriminated to (high-frequency) oscillating modes and (low-frequency) non-oscillating ones. The {\em separatrix} between these modes is set at $k_c = \sqrt{\frac{8 \pi G}{3 c^2} M_2}$, where $M_2$ is the initial amount of the linear defects, corresponding to a present-time frequency, $f_c$, given by Eq. (56) of~\cite{1}.

\item Due to quantum-gravitational effects, the {\em high-frequency part} $(k > k_c)$ of the corresponding CGW signal is being further {\em amplified} by a factor of $100$ (see, e.g., Fig. 1), while, its {\em profile} is turned to a {\em periodic function} of the frequency, the period of which depends (solely) on the duration $(\Dl \ta)$ of the radiation-plus-strings stage. To the best of our knowledge, such an amplification mechanism of a stochastic GW signal has not yet been considered elsewhere. 

\item According to our model, the only interference between an early-radiation epoch and the late-radiation era comes from a period of significant presence of cosmic strings (encapsulated in the cosmological model). Therefore, it is natural to assume that any {\em energization} of the gravitational-perturbation field (graviton production) is due to a corresponding {\em energy-loss} of the long-linear-defects network. In this context, it is possible that the actual {\em scaling process} might have been completed a little bit earlier than what is currently anticipated by the scientific community (see, e.g.,~\cite{29},~\cite{30},~\cite{32}). 

\end{itemize}

As we have already mentioned, the number of gravitons, $N_{k_c}$, created in the {\em ground state} of the oscillating metric perturbations, is well-defined and finite. These modes enter inside the horizon at some (coordinate) time $t_c > t_{sc}$, at which, their {\em physical wavelength} (defined as $\lm_{ph} (t_c) = \lm_c \: R (t_c)$) becomes smaller than the corresponding {\em Hubble radius} ($\ell_H (t_c) = c / H (t_c)$, $H$ being the Hubble parameter). According to~\cite{1}, we have \be t_c \: \gtrsim \: \frac{3 \pi}{11 \gm_*^2} \: \left ( \frac{G \mu}{c^2} \right )^{-2} t_{cr} , \ee where $t_{cr}$ is the time at which the Universe acquires the {\em critical temperature} below which the cosmic strings are formed (for GUT-scale strings, $t_{cr} \sim 10^{-31} \;sec$) and $\gm_*$ is the number of correlation lengths $(4 \leq \gm_* \leq 7)$ inside the horizon at $t_* = (\frac{G \mu}{c^2})^{-1} t_{cr}$ (see, e.g.,~\cite{7}). 

The dimensionless quantity $(G \mu / c^2)$ characterizes the strength of the gravitational interactions of strings. The observations give an upper limit on the value of this parameter. In particular, the current CMB bound is~\cite{37} $(G \mu / c^2) \leq 1.3 \times 10^{-6}$ while, more recent studies on a cosmic-string contribution in the WMAP data~\cite{38} have yielded the tighter bound $(G \mu / c^2) \leq 3.3 \times 10^{-7}$, something that is also confirmed by gravitational lensing observations: $(G \mu / c^2) \geq 4 \times 10^{-7}$~\cite{39}, \cite{40}. Therefore, in what follows, we may consider $(G \mu / c^2) = \frac{1}{\gm_*} \times 10^{-6}$, so that, $t_c \gtrsim 10^{12} \: t_{cr} \sim 10^{-19} \; sec$. 

As far as the oscillating modes are concerned, $k_c$ is the lowest-allowed comoving wave-number. Therefore, after $t_c$, no oscillating CGW modes (created by this mechanism) are left, to enter inside the Hubble sphere. For $t > t_c$, the only contribution to the CGW-energy density comes from the {\em low-frequency}, non-oscillatory modes, of comoving wave-number $k < k_c$.

\section{The extra GW-energy density}

Now, we may proceed even further, to calculate the (integrated) extra GW-energy density present in the late-radiation epoch, due to the metric perturbations created in the scaling of the long-cosmic-string network. According to what previously stated, two kinds of modes are present in the late-radiation era. We consider each case separately:

\subsection{Oscillating modes}

The integrated energy density of the oscillating modes $(k > k_c)$, created in the transition of the Universe through the radiation-plus-strings stage, is given by \be {\cal E}_{gw} (k > k_c) = \int_{k_c}^{\infty} d \eps_{k > k_c} (k) = \int_1^{\infty} d \eps_{k > k_c} (x) , \ee which, in view of Eqs. (38) and (55), is written in the form \bea {\cal E}_{gw} (k > k_c) & = & \pi \: \eps_c (\ta) \frac{1}{N_{k_c}} \\ & \times & \int_1^{\infty} \frac{\sin^2 \left ( 2 \sqrt{N_{k_c}} \: \sqrt{x^2 - 1} \right )}{x^2 - 1} \: x dx \nn \eea or, else, \bea && {\cal E}_{gw} (k > k_c) = 4 \pi \: \eps_c (\ta) \: \frac{1}{N_{k_c}} \times \\ && \int_1^{\infty} \frac{\sin^2 \left ( \sqrt{N_{k_c}} \: \sqrt{x^2 - 1} \right ) \: \cos^2 \left ( \sqrt{N_{k_c}} \: \sqrt{x^2 - 1} \right )}{x^2 - 1} x dx . \nn \eea In fact, the high-frequency gravitons, responsible for the infinite upper-limit, can be produced only by an instantaneous change in the scalar curvature. In our case, the {\em highest frequency} (which can be produced) is determined by the speed of the transition (the duration of the radiation-plus-strings stage), otherwise particle creation is {\em adiabatically suppressed} (see, e.g.,~\cite{20},~\cite{21},~\cite{34}). Hence, \be \gm k c = c \sqrt {k^2 - k_c^2} \leq \frac{1}{\Dl \ta} \ee (cf. Eqs. (21) and (22)) and, therefore, \be \ch = \sqrt {x^2 - 1} \leq \frac{1}{2 \sqrt{N_{k_c}}} . \ee Now, in terms of the parameter $\ch$, Eq. (63) is written in the form \bea &&{\cal E}_{gw} (k > k_c) = 4 \pi \: \eps_c (\ta) \: \frac{1}{N_{k_c}} \times \\ && \int_0^{\frac{1}{2\sqrt{N_{k_c}}}} \; \frac{\sin^2 \left ( \sqrt{N_{k_c}} \: \ch \right ) \: \cos^2 \left ( \sqrt{N_{k_c}} \: \ch \right )}{\ch} d \ch \nn \eea and the infinite upper-limit can be recovered under the assumption $N_{k_c} \ll 1$ (in fact, this assumption is quite accurate, since, according to~\cite{1}, $N_{k_c} = 2 \times 10^{-3}$). However, in this case, we have \bea \sin \left ( \sqrt{N_{k_c}} \: \ch \right ) & = & N_{k_c}^{1/2} \ch - \frac{1}{6} N_{k_c}^{3/2} \ch^3 + \textsl{O} \left ( N_{k_c}^{5/2} \right ) \nn \\ \cos \left ( \sqrt{N_{k_c}} \: \ch \right ) & = & 1 - \frac{1}{2} N_{k_c} \ch^2 + \textsl{O} \left ( N_{k_c}^2 \right ) \: . \eea Accordingly, discarding terms of order $\textsl{O} \left ( N_{k_c}^2 \right )$, Eq. (66) is written in the form \be {\cal E}_{gw} (k > k_c) \simeq 4 \pi \: \eps_c (\ta)  \int_0^{\frac{1}{2\sqrt{N_{k_c}}}} \; \left ( \ch - \frac{4}{3} N_{k_c} \ch^3 \right ) d \ch, \ee thus, resulting in \be {\cal E}_{gw} (k > k_c) \simeq \frac{5 \pi}{12} \: \eps_c (\ta)\: \frac{1}{N_{k_c}} . \ee By virtue of Eq. (54), Eq. (69) takes on its final form \be {\cal E}_{gw} (k > k_c) \simeq \frac{1}{R^4 (t)} \: \frac{5 \hbar c}{48 \pi^2} \: k_c^4 . \ee

Inside the Hubble sphere of the late-radiation era, the energy density of the gravitons with $k > k_c$ which were created at earlier times $t_{sc} \leq t^{\prime} < t$ and the wavelength of which became less than the horizon due to the expansion of the Universe at $t_{sc} \leq t \leq t_c$, is given by the sum of all the previous increments $\dl {\cal E}_{gw} (t^{\prime})$ being redshifted by the factor $R^4 (t^{\prime}) / R^4 (t)$ \bea \varep_{gw} \left ( t_{sc} \leq t \leq t_c \right ) & = & \int_{t_{sc}}^{t_c} \frac{R^4 (t^{\prime})}{R^4 (t)} \left ( - \frac{d}{d t^{\prime}} {\cal E}_{gw} \right ) d t^{\prime} \Rarrow \\ \varep_{gw} \left ( t_{sc} \leq t \leq t_c \right ) & = & \frac{5 \hbar c}{24 \pi^2} \: k_{c_{ph}}^4 (t_{sc}) \left ( \frac{t_{sc}}{t_c} \right )^2 \ln \left ( \frac{t_c}{t_{sc}} \right ). \nn \eea For $t > t_c$ (i.e., deep inside the late-radiation era), this constant amount of energy density is being (further) redshifted due to the cosmological expansion. Therefore, \bea \varep_{gw}^{k > k_c} (t) & = & \varep_{gw} \left ( t_{sc} \leq t \leq t_c \right ) \frac{R^4 (t_c)}{R^4 (t)} \Rarrow \nn \\ \varep_{gw}^{k > k_c} (t) & = & \frac{5 \hbar c}{24 \pi^2} \: k_{c_{ph}}^4 (t_c) \left ( \frac{t_c}{t} \right )^2 \ln \left ( \frac{t_c}{t_{sc}} \right ), \eea while, for $t \geq t_c$, there are only the non-oscillatory modes available to enter inside the horizon and contribute to the extra GW-energy density.

\subsection{Non-oscillating modes}

The extra GW-energy density (present in the late-radiation era) of the modes with $k < k_c$, created in the transition of the Universe through the radiation-plus-strings stage, is given by \be {\cal E}_{gw} (k < k_c) = \int_0^{k_c} d \eps_{k < k_c} (k) = \int_0^1 d \eps_{k < k_c} (x) , \ee which, by virtue of Eqs. (38) and (58), is written in the form \bea {\cal E}_{gw} (k < k_c) & = & \pi \: \eps_c (\ta) \frac{1}{N_{k_c}} \\ & \times & \int_0^1 \frac{\sinh^2 \left ( 2 \sqrt{N_{k_c}} \sqrt{1 - x^2} \right )}{1 - x^2} \: x \: dx \: . \nn \eea Now, in terms of the parameter $\ph = \sqrt{1 - x^2}$, Eq. (74) reads \bea {\cal E}_{gw} (k < k_c) & = & 4 \pi \: \eps_c (\ta) \: \frac{1}{N_{k_c}} \\ & \times & \int_0^1 \frac{\sinh^2 \left ( \sqrt{N_{k_c}} \: \ph \right ) \: \cosh^2 \left ( \sqrt{N_{k_c}} \: \ph \right )}{\ph} d \ph , \nn \eea which, in the limit $N_{k_c} \ll 1$, results in \be {\cal E}_{gw} (k < k_c) \simeq 4 \pi \: \eps_c (\ta) \int_0^1 \left ( \ph + \frac{4}{3} N_{k_c} \ph^3 \right ) d \ph , \ee yielding \be {\cal E}_{gw} (k < k_c) \simeq 2 \pi \: \eps_c (\ta) \left ( 1 + \frac{2}{3} N_{k_c} \right ), \ee or, else, \be {\cal E}_{gw} (k < k_c) \simeq \frac{1}{R^4 (t)} \: N_{k_c} \frac{\hbar c}{2 \pi^2} \: k_c^4 . \ee For the same reasons that led to Eq. (71), the energy density of the gravitons with $k < k_c$ which were created at earlier times $t_c \leq t^{\prime} < t$ and the wavelength of which became less than the horizon due to the expansion of the Universe at $t_c \leq t \leq t_{rec}$ (where $t_{rec}$ is the {\em recombination time}), is given by \be \varep_{gw}^{k < k_c} (t) = N_{k_c} \frac{\hbar c}{\pi^2} \: k_{c_{ph}}^4 (t_c) \left ( \frac{t_c}{t} \right )^2 \ln \left ( \frac{t}{t_c} \right ). \ee 

\subsection{The overall energy density}

With the aid of Eqs. (72) and (79), we can determine the (overall) extra GW-energy density, due to the metric perturbations created in the scaling of a long-cosmic-string network and entered inside the horizon at $t_c \leq t \leq t_{rec}$. We obtain \bea \varep_{gw}^{rps} (t \geq t_c) & = & \frac{\hbar c}{\pi^2} \: k_{c_{ph}}^4 (t_c) \left ( \frac{t_c}{t} \right )^2 \\ & \times & \left [ \underbrace{\frac{5}{24} \ln \left ( \frac{t_c}{t_{sc}} \right )}_{oscillating} + \underbrace{N_{k_c} \ln \left ( \frac{t}{t_c} \right )}_{non-oscillating} \right ], \nn \eea where we have marked the contribution of both the oscillating and the non-oscillating modes. By definition, the physical wavelength of the mode $k_c$ at $t_c$ equals to the horizon length (at $t_c$ this mode enters into the visible Universe), so that \be k_{c_{ph}} (t_c) = \frac{2 \pi}{\lm_{c_{ph}} (t_c)} = \frac{2 \pi}{\ell_H (t_c)} = \frac{2 \pi H (t_c)}{c} . \ee Accordingly, Eq. (80) is written in the form \bea \varep_{gw}^{rps} (t \geq t_c) & = & 16 \pi^2 \frac{\hbar}{c^3} \: H^4 (t_c) \left ( \frac{t_c}{t} \right )^2 \\ & \times & \left [ \frac{5}{24} \ln \left ( \frac{t_c}{t_{sc}} \right ) + N_{k_c} \ln \left ( \frac{t}{t_c} \right ) \right ] , \nn \eea which, in view of the Appendix A, results in \bea \frac{\varep_{gw}^{rps}}{\varep_{gw}^{inf}} (t \geq t_c) & = & \frac{247808}{81} \pi^2 \gm_*^4 \left ( \frac{G \mu}{c^2} \right )^4 \left [ \frac{H_{cr}}{H_{inf}} \right ]^2 \left ( \frac{t_0}{t} \right )^2 \nn \\ & \times & \left [ \frac{5}{24} \ln \left ( \frac{t_c}{t_{sc}} \right ) + N_{k_c} \ln \left ( \frac{t}{t_c} \right ) \right ], \eea where $\varep_{gw}^{inf}$ is the GW-energy density predicted by (de Sitter-type) inflation, $H_{inf}$ and $H_{cr}$ are the values of the Hubble parameter at inflation and at $t_{cr}$, respectively, while $t_0 \simeq 4.3 \times 10^{17} \; sec$ denotes the present epoch. According to Allen~\cite{20}, during the inflationary epoch, $H_{inf} = 8 \times 10^{34} \; sec^{-1}$, while, within the late-radiation era, $H_{cr} = \left (2 t_{cr} \right )^{-1} = 5 \times 10^{30} \; sec^{-1}$. Finally, admitting that $\Dl \ta = 5.5 \: \ta_*$ (see, e.g.,~\cite{1}), we obtain $N_{k_c} = 2 \times 10^{-3}$, $t_{sc} \simeq 30 \: t_*$ and, therefore, $t_c \simeq 10^4 \: t_{sc}$. Inserting these values into Eq. (83) we obtain \be \frac{\varep_{gw}^{rps}}{\varep_{gw}^{inf}} (t \geq t_c) = 4.28 \times 10^7 \left ( \frac{sec}{t} \right )^2 \left [ 1 +  10^{-3} \ln \left ( \frac{t}{sec} \right ) \right ]. \ee We observe that, at late times, the only contribution to $\varep_{gw}^{rps}$ comes from the second term in the brackets on the {\em right-hand-side} of Eq. (84), i.e., from the (low-frequency) non-oscillatory CGW modes (a not unexpected result). Nevertheless, within the late-radiation era, this term introduces only minor corrections to the $\frac{1}{t^2}$ behavior of the CGW-energy density. 

It is generally admitted that the main source of relic gravitational radiation is the inflationary amplification of metric perturbations, taken place as early as $10^{-34} \; sec$ after the Big Bang (see, e.g.,~\cite{12}). Quite later, $t \geq t_{sc} \sim 10^{-23} \; sec$, after being {\em filtered} by a long-cosmic-string network, the original number of the relic (inflationary) GWs is further increased by an additional amount of CGWs (cf. Eqs. (36) and (43)), due to the quantum-gravitational creation of gravitons along the transition of the Universe through the radiation-plus-strings stage. In this case, Eq. (84) indicates something very interesting:  At times $t_c \sim 10^{-19} \; sec \leq t \leq 6540 \; sec$, the contribution of these gravitons, $\varep_{gw}^{rps}$, to the (already) existing gravitational-energy density $\varep_{gw}^{inf}$, is (much) larger than $\varep_{gw}^{inf}$ itself. 

Certainly, at late times, the amount, $\varep_{gw}^{rps}$, of the {\em extra} gravitational-energy density (created in the scaling of a long-cosmic-string network) becomes negligible, as a consequence of the cosmological expansion. Nevertheless, it may have affected significantly the cosmological evolution within the first (and most interesting) $109 \; min$ of the history of the Universe, thus leaving imprints on the anticipated stochastic GW power spectrum. Accordingly, for $f \geq f_c$ (see, e.g.,~\cite{1}), the spectral intensity $\Om_{gw}$ of the CGW signal is turned to a periodic function of the frequency, while being amplified by (almost) two orders of magnitude (Fig. 1).

\section{Conclusions}

Assuming that there was a period in the early Universe in which the matter content could be modelled by a two-component fluid (consisting of radiation and a cosmic-string fluid), we demonstrate that the existence of such a (radiation-plus-strings) stage would have resulted in a measurable effect on the CGWs' power spectrum. In fact, along the transition from an early-radiation epoch to the late-radiation era through the radiation-plus-strings stage, {\em gravitons} with a highly-recognizable profile could have been produced. 

As a consequence, at high frequencies, the CGW spectrum would no longer be {\em scale invariant}, but, rather, a {\em periodic} function of the frequency, the period of which depends (solely) on the duration of the radiation-plus-strings stage.

But, what is more important for the detection of CGWs, is that, the extra amount of metric perturbations created in the scaling of a long-cosmic-string network would lead to the {\em amplification of the relic GW signal} by (almost) two orders of magnitude. 

The {\em extra-energy} density of the CGWs created in the transition from an early-radiation epoch to the late-radiation era through the radiation-plus-strings stage, not only is added to the corresponding inflationary quantity. In fact, at times $10^{-19} \; sec \leq t \leq 6540 \; sec$, i.e., within the first (and, certainly, most interesting) $109 \; min$ of the Universal evolution, dominates over every other form of GW-energy density.

According to our model, the only interference between an early-radiation epoch and the late-radiation era comes from a period of a significant presence of cosmic strings (encapsulated in the cosmological model). Therefore, it is natural to assume that any {\em energization} of the gravitational-perturbation field (graviton production) is due to a corresponding {\em energy-loss} of the long-linear-defects network. In this article we have neglected the backreaction of the created gravitons on the curved cosmological background (it will be the scope of a future work); nevertheless, we expect that the actual {\em scaling process} might have been completed a little bit earlier than what is currently anticipated by the scientific community. 

\section*{Acknowledgements} 

The first author (K. K.) would like to thank the Research Committe of the Technological Education Institute of Serres, for the finacial support under the grant SAT/ME/210911-100/03, and Enric Verdaguer acknowledges support by the Spanish Research Projects MEC FPA-2007-66665, CPAN CSD-2007-00042, and AGAUR 2009-SGR-00168.

\section*{Appendix A}

The contribution of the gravitons created in the transition from an early-radiation epoch to the late-radiation era through a radiation-plus-strings stage to the total amount of the gravitational-energy density in the Universe, is given by $$ \varep_{gw}^{rps} (t \geq t_c) = 16 \pi^2 \frac{\hbar}{c^3} \: H^4 (t_c) \left ( \frac{t_c}{t} \right )^2 $$ \\ $$\times \left [ \frac{5}{24} \ln \left ( \frac{t_c}{t_{sc}} \right ) + N_{k_c} \ln \left ( \frac{t}{t_c} \right ) \right ] \eqno{(A1)} $$ (cf. Eq. (82)), which, upon consideration of the Friedmann equation $$ H^2 (t_c) = \frac{8 \pi G}{3 c^2} \varep_{rad} (t_c), \eqno{(A2)}$$ reads $$\varep_{gw}^{rps} (t \geq t_c) = \frac{128 \pi^3}{3} H^2 (t_c) \: t_{Pl}^2 \: \varep_{rad} (t) $$ \\ $$\times \left [ \frac{5}{24} \ln \left ( \frac{t_c}{t_{sc}} \right ) + N_{k_c} \ln \left ( \frac{t}{t_c} \right ) \right ], \eqno{(A3)} $$ where $t_{Pl} = 5.44 \times 10^{-44} \; sec$ is the Planck time. Taking into account that, in terms of $\varep_{rad} (t_c)$, the mass-energy density in the matter-dominated era is given by $$ \varep_{matt} (t) = \varep_{rad} (t_c) \left ( \frac{t_c}{t_0} \right )^2 , \eqno{(A4)} $$ where $t_0$ denotes the present epoch, we find that $$\varep_{gw}^{rps} (t \geq t_c) = \frac{128 \pi^3}{3} H^2 (t_c) \: t_{Pl}^2 \: \varep_{matt} (t) \left ( \frac{t_0}{t} \right )^2 $$ \\ $$\times \left [ \frac{5}{24} \ln \left ( \frac{t_c}{t_{sc}} \right ) + N_{k_c} \ln \left ( \frac{t}{t_c} \right ) \right ], \eqno{(A5)} $$ which, upon definition of $H_{inf}$ as the Hubble parameter at (de Sitter) inflation, can be written in the more convenient form $$\varep_{gw}^{rps} (t \geq t_c) = \frac{3}{16 \pi} H_{inf}^2 t_{Pl}^2 \frac{2048}{9} \pi^4 \left [ \frac{H (t_c)}{H_{inf}} \right ]^2 \varep_{matt} (t) \left ( \frac{t_0}{t} \right )^2 $$ \\ $$\times \left [ \frac{5}{24} \ln \left ( \frac{t_c}{t_{sc}} \right ) + N_{k_c} \ln \left ( \frac{t}{t_c} \right ) \right ] \eqno{(A6)} $$ and, by virtue of Eq. (4.10) of~\cite{20} (or Eq. (16) of~\cite{34}), takes on its final form $$\varep_{gw}^{rps} (t \geq t_c) = \varep_{gw}^{inf} \: \frac{2048}{9} \pi^4 \left [ \frac{H (t_c)}{H_{inf}} \right ]^2 \left ( \frac{t_0}{t} \right )^2 $$ \\ $$\times \left [ \frac{5}{24} \ln \left ( \frac{t_c}{t_{sc}} \right ) + N_{k_c} \ln \left ( \frac{t}{t_c} \right ) \right ] . \eqno{(A7)} $$ Eventually, taking into account that, in a FRW radiation model we have $\frac{H(t_c)}{H_{cr}} = \frac{t_{cr}}{t_c}$, with the aid of Eq. (60), Eq. (A7) results in Eq. (83).

\end{document}